\documentclass[11pt]{article}
\usepackage{graphicx}
\usepackage[margin=1in]{geometry}

\usepackage{makeidx}  % allows for indexgeneration
\usepackage{ifpdf}
\usepackage{url}

\title{Elementary Cellular Automata as Multiplicative Automata}
\date{February 2025}
\author{Daniel McKinley}

\begin{document}

\maketitle

\section{Introduction}

Elementary cellular automata (ECA) are a set of simple binary programs in the form of truth tables called Wolfram codes that produce complex output when done repeatedly in parallel, and quaternions are an extension of complex numbers frequently used to represent 3D space and its rotations in computer graphics due to the lack of gimbal lock present with Euler angles. Both are well-studied subjects, this Java library puts them together in a new way. This project changes classical additive cellular automata into multiplicative automata \cite{Wolfram} via permutations and hypercomplex numbers as binary pointer arrays. Valid solutions extend the binary ECA to complex numbers, produce a vector field, make an algebraic polynomial, and are verified in several ways. The Java code implementation \cite{mygit} and a solution image database \cite{dmwebsite} is available at the GitHub and website in the references.

Very loosely analogous to De Morgan's law in Boolean algebra, the main algorithm produces several multiplicative versions of any given standard additive binary Wolfram code up to 32 bits and is written to support user supplied complex 1-D input at row 0 with choice of type of multiplication tables and partial product tables among other parameters. It produces an algebraic polynomial and complex vector field output for any given Wolfram code, and the hypercomplex 5-factor identity solution allows for the complex extension of any binary cellular automata.  The automorphism libraries may be of value to the open source community as well. 

There are other cellular automata implementations, Mathematica \cite{Mathematica}, CellPyLib \cite{Antunes2021}, and others. This is not designed to replace those general purpose utilities, it's focused on the set of Wolfram code operation conversions. The GUI is designed to show enough to conclude that the math works and give a rough idea of aggregate behavior over parameters and the algorithm code is designed to split off and plug in somewhere else. There are some useful things you can build on it directly or indirectly, such as making Bloch spheres out of layers of complex number output, enabling an extension of Fourier analysis, a complex version of the prime number automata \cite{Wolfram}, that are clear directions to go in but subject to a different set of decisions like application-specific tech debt and potential translation to C++ or Python and out of scope of this paper.

\section{Automorphisms}
\begin{center}
Hypercomplex unit vector implementation
\includegraphics{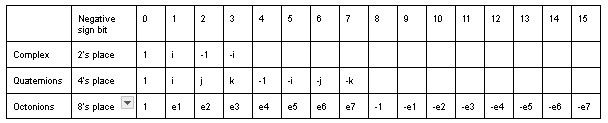}\\
\end{center}

The Cayley-Dickson (CD) and Fano automorphism support classes are discussed in greater detail in the readme and the documentation, they along with the Galois class provide sets of multiplication tables to be validated with Wolfram codes and used in output computation. The CD multiplication implementation permutes the steps of splitting and recombining hypercomplex numbers to increase the scope of the CD equation, $(a,b)x(c,d)=(ac-d*b,da+bc*)$, where * is the conjugate. It verifies itself by producing the vector product of two symmetric groups of its degree operating on the up/down recursion factoradics when interacting with other CD multiplications, and is the permutations of the unit vector bit layers excluding the negative sign.  It is equivalent to selecting all vertexes by sequences of faces of a hypercube.

\section{Main Algorithm}

The main algorithm uses a set of permutations operating on cellular automata input, each permutation permuting the neighborhood, becoming a factor, with four kinds of multiplications. The multiplication tables are input as 2D but used as N-D, where N=numFactors.
\begin{quote}
\subsection{Multiplications A, additive to multiplicative}
\begin{itemize}
\item r = specific Wolfram code
\item n = neighborhood sum = $1*columnZero + 2*columnOne+ 4*columnTwo...2^(columnN)*columnCol$, points to its value in r
\item h = hypercomplex unit vector from binary
\item H = inverse of h, binary value from hypercomplex unit vector
\item p = a permutation of the neighborhood
\item Using hypercomplex multiplication, a valid permutation set produces:
\item $WolframCode(r, n) = WolframCode(r,  H(h(p(n)) * h(p(n)) * h(p(n)) ... numFactors)$, though n may or may not equal H(...)
\item $WolframCode(r, H(h(p(n)) * h(p(n)) * h(p(n)) ... numFactors))$ is a pointer array that always points to an equal value $(0,1)$ within $WolframCode(r, _)$
\item Each $h(p(n))$ in a valid solution is a factor template in the multiplication table for all values of its axis
\end{itemize}
\end{quote}
\begin{center}
\includegraphics{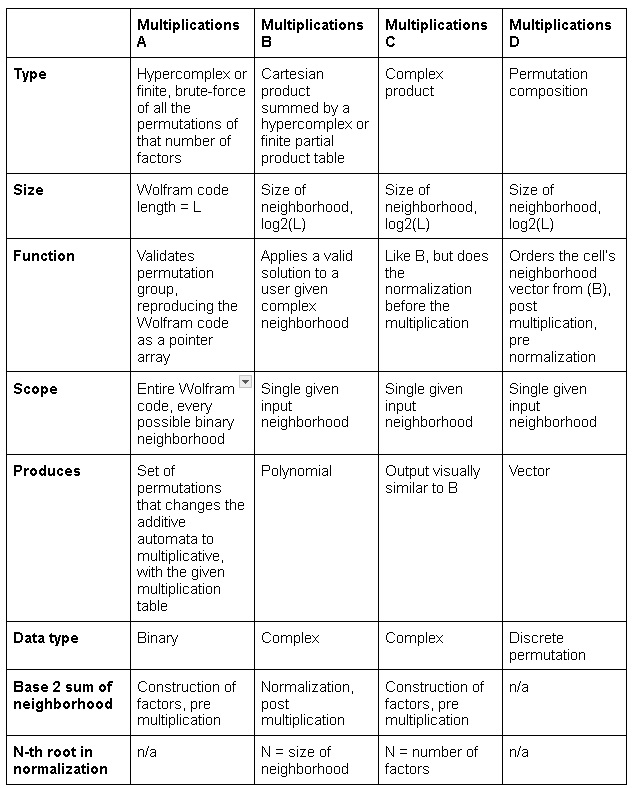}
\end{center}
\pagebreak

The first set of multiplications, column A, brute forces all possible sets of permutations on all possible binary neighborhoods of the Wolfram code. A permutation in the set rearranges the columns of the input neighborhood (not the multiplication table columns themselves, the bits addressing the columns) for each layer, these become a set of factors.  A valid set of permutations is one that, for all possible input neighborhoods, the set of constructed factors using the permuted neighborhoods always multiplies out to a value that points to an equal value within the Wolfram code. The set of multiplication results is a pointer array that reproduces the original Wolfram code for every possible binary neighborhood. 

Identity solutions of $5+4n$ factors using all zero permutations exist for Wolfram codes up to 32 bits in this library using hypercomplex numbers. The factors constructed are a loose diagonal through the multidimensional multiplication table, starting at the origin and ending at the opposite corner while zig-zagging. The path lengths of each factor and the result are included in ValidSolution results.

\begin{center}
Permutations of 3 bit neighborhoods\\
\includegraphics{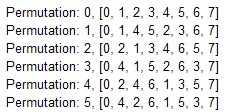}

Flattened path through a six dimensional multiplication table
Six factors, permutation set = {0,1,2,3,4,5}\\
\includegraphics{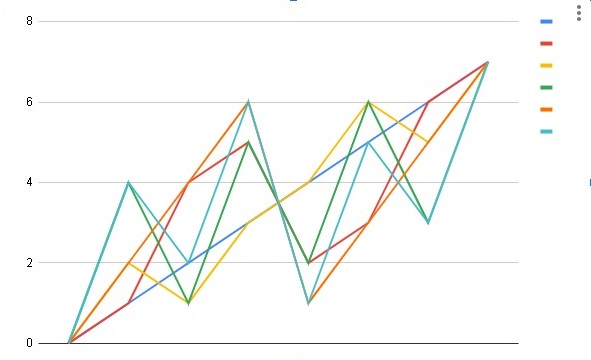}
\end{center}

Multiplications B and C apply a valid solution from the first set of multiplications to any given individual neighborhood with binary, non-negative real, and complex values. Multiplication B is the Cartesian product of the permuted neighborhoods, using a closed partial product table to generate a polynomial. Multiplication C does the binary sum of complex neighborhood, then multiplies as complex. Both B and C take the n-th root of the result, with n = numColumns and n = numFactors, respectively. Multiplications B and C both include a binary weighted sum of the neighborhood, same as the construction of the factors from A, though B and C use complex. B, as part of the normalization and C as the construction. Multiplication C is the permutation composition product. B, just before the normalization is a neighborhood of multiplication results, with each column of it being a unit vector coefficient. This multiplication result neighborhood is permuted by the inverse of the permutation composition product to properly order the output vector. There are a couple of normalization parameters and a hybrid multiplicative-additive output option that are discussed more in the documentation.

\section{Code tests}

There are several algorithms implemented that verify the integrity of the permutations, automorphisms, polynomial, and graphic output. For all degrees, the Cayley-Dickson automorphism interaction produce two layers of the symmetric group $((cdz,cdo,cdzz,cdoo))) = (row0,row1),(row1,row0) = places$ symmetric group layer independently of the d!'s place layer. The Fano automorphisms' first set of triplets is equal to the CD hypercomplex numbers The coefficients of the polynomials produced always sum to $neighborhoodSize^numFactors$. The graphed output for a function is on two panels with two seperate sets of multiplications; one multiplication uses the entire partial product table expansion, the other uses the polynomial, and these two panels output the exact same. Every automorphism used can be tested for 2D Latin-ness, which in itself does not prove complete correctness, but provides an easy first-pass filter. The Galois fields produce Mutually Orthogonal Latin Squares (MOLS) in sets of $(p^m)-1$, which are easily verified 
\begin{center}

\includegraphics{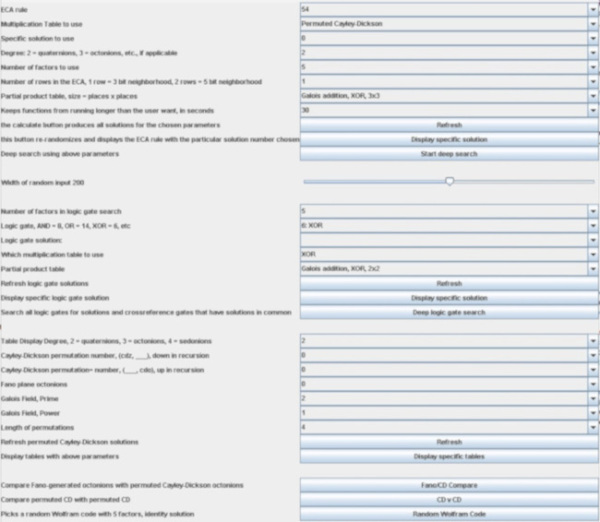}
Control Panel

\includegraphics{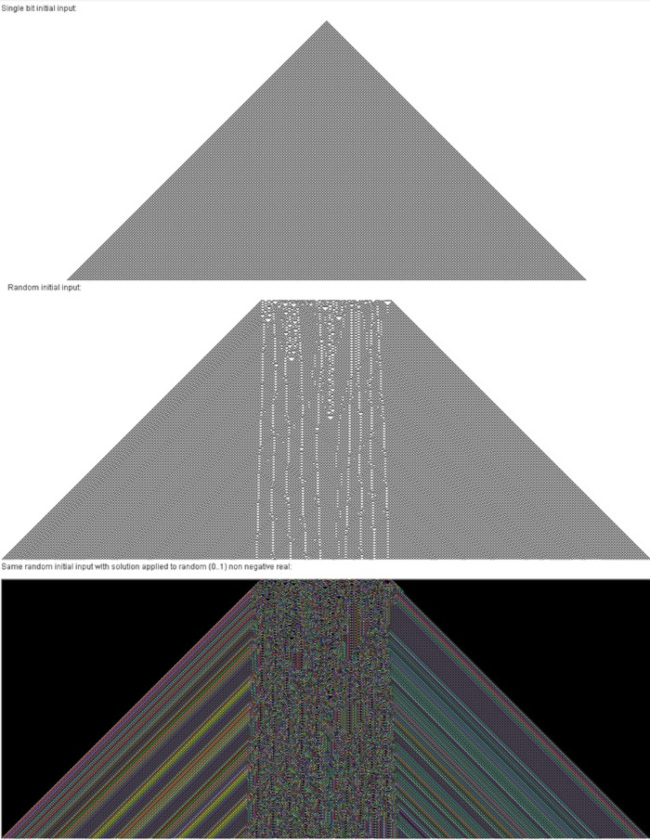}
ECA 54, binary and non-negative real

\includegraphics{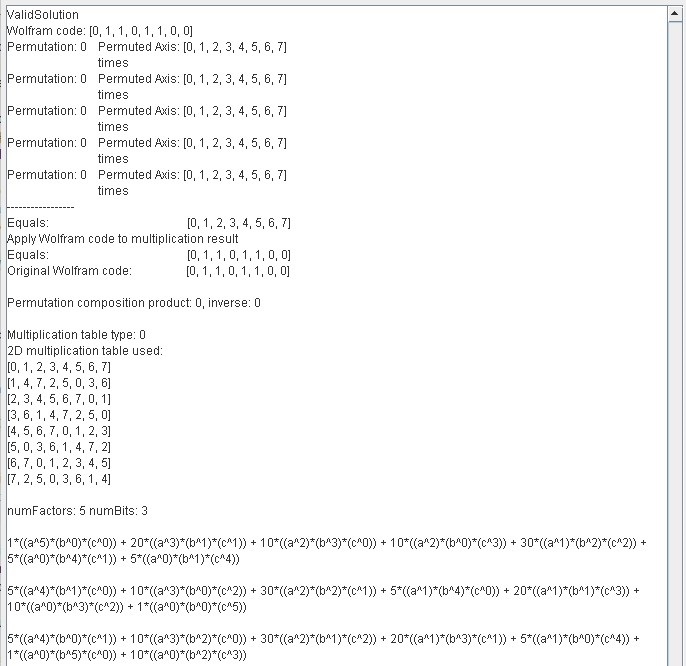}
ECA 54, solution parameters, including polynomial

\includegraphics{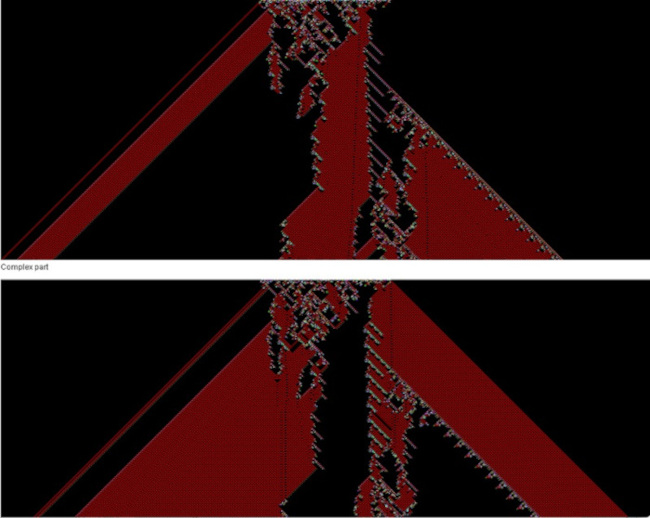}
ECA 54, solution output, complex
\end{center}
\medskip
\bibliographystyle{plain}
\bibliography{MultiplicativeECA.bbl}

\end{document}